\documentclass{blois}
\pdfoutput=1
\title{A new approach to achieving high granularity for silicon diode detectors with impact ionization gain}

\usepackage{HGTD_TDR-defs}
\usepackage{multirow}
\usepackage{rotating}
\usepackage{mathrsfs}
\usepackage{booktabs}
\usepackage{longtable}
\usepackage{placeins}
\usepackage{amssymb}
\usepackage{upgreek}
\usepackage{svg}
\usepackage{comment}
\usepackage{pdfpages}
\usepackage{siunitx}
\usepackage{lineno}
\usepackage{float}

\usepackage{cleveref}
\crefname{equation}{Eq.}{Eqs.}
\crefname{figure}{Fig.}{Figs.}
\crefname{table}{Tab.}{Tabs.}
\crefname{section}{Sec.}{Secs.}

\newcommand*{\ATLASLATEXPATH}{latex/}
\usepackage{\ATLASLATEXPATH atlasphysics}
\usepackage{\ATLASLATEXPATH atlasxref}

\newcolumntype{P}[1]{>{\raggedright\arraybackslash}p{#1}}

\begin{document}
\maketitle
{\small \centering
            S.~Ayyoub$^b$, C.~Gee$^a$, R.~Islam$^b$, S.~M.~Mazza$^{a}$, \\ B.~Schumm$^a$, A.~Seiden$^a$, Y.~Zhao$^a$\\
}
\bigskip
{\small \centering \it
            $^a$The Santa Cruz Institute for Particle Physics and the University of California, Santa Cruz, California, 95064\\
            $^b$CACTUS Materials, Inc., Tempe Arizona, 85284\\
}

\begin{abstract}
Low Gain Avalanche Diodes (LGADs) are thin (20-50 $\mu m$) 
silicon diode sensors with modest internal gain (typically 5 to 50) and exceptional time resolution (17~$ps$ to 50~$ps$).
However, the granularity of such devices is limited to the millimeter scale due to the need to include protection structures at the boundaries of the readout pads to avoid premature breakdown due to large local electric fields. In this paper we present a new approach -- the Deep-Junction LGAD (DJ-LGAD) -- that decouples the high-field gain region from the readout plane. This approach is expected to improve the achievable LGAD granularity to the tens-of-micron scale while maintaining direct charge collection on the segmented electrodes.
\end{abstract}

\section{Introduction}

Low Gain Avalanche Diodes (LGADs)~\cite{bib:LGAD,bib:MarTorino,bib:UFSD300umTB} are a type of thin silicon diode sensor, typically implemented with an n-on-p architecture with the addition of a highly doped p+ layer just below the n-type implants of the electrodes. 
This additional layer, called the ``gain" or ``multiplication" layer, generates a high field region where controlled charge multiplication is possible, and can be up to a few microns thick. The remainder of the sensor, which is typically composed of high-resistivity silicon, is referred to as the ``bulk".
Owing to the intrinsic impact-ionization gain (which is typically between 5 and 50) these devices can be very thin (20-50~$\mu m$) while achieving charge collection levels greater than their much thicker conventional counterparts. This allows a short collection time with a fast rising edge that results in very precise timing information (17-50~$ps$) \cite{Padilla_2020, JIN2020164611, Mazza:2018jiz, bib:HPKirradiation35vs50, bib:HPKirradiationGalloway}.
LGADs were first developed by the Centro Nacional de Microelectrónica (CNM) Barcelona, with significant participation by the RD50 Collaboration \cite{rd50}.
Due to their precise timing capability, LGADs offer a prospective new paradigm for space-time particle tracking~\cite{bib:4Dtracking}. 

The first application of LGADs is planned for the High Luminosity LHC (HL-LHC~\cite{bib:hllhc}), where extreme beam-collision pileup conditions will lower the efficiency for tracking and vertexing for the inner tracking detector in the region close to the beam pipe. To restore this performance, LGAD-based timing layers that can time-stamp collision vertices are being developed for the forward region of both the ATLAS and the CMS experiments. 
The ATLAS and CMS projects are called, respectively, the High Granularity Timing Detector (HGTD)~\cite{CERN-LHCC-2020-007} and the End-Cap Timing Layer (ETL)~\cite{CMS:2667167}. The pixellated LGADs that will be used in these timing layers will make use of an inter-channel protection structure referred to as the ``Junction Termination Extension" (JTE). These structures, which avoid breakdown between adjacent readout channels by terminating the high field in the gain layer (see~\cref{fig:jte}), create ``dead regions" of order 50-100~$\mu m$ between channels, in which the charge collection is severely limited. As a result, the granularity of the LGAD sensors under development for use at the HL-LHC is limited to the millimeter scale. Although this degree of granularity is acceptable for HL-LHC applications, many possible future applications of LGADs, such as four-dimensional tracking at future colliding beam facilities, X-ray imaging, and medical physics applications, will require granularity at the tens-of-micron scale~\cite{ref:Jeph_Wang}.

In this paper an innovative LGAD design, referred to as the ``Deep Junction LGAD" (DJ-LGAD), will be presented. The Deep Junction approach permits granularity on the same scale as that of conventional silicon diode sensors, while maintaining a direct coupling of the signal charge to the readout electrodes. This new design features a multiplication zone that is de-coupled from the readout plane by burying a high-field diode junction several microns below the surface of the device, separated from the surface readout plane by a region of lower field that is still high enough to maintain drift-velocity saturation. In this way the high field area is kept sufficiently far from the segmented area of the silicon so that inter-channel breakdown is avoided, allowing for the standard pixelization of the readout plane, and the achievement of granularity as fine as tens of microns.

\section{Background and Motivation}

LGADs were first proposed as an approach to achieving precise timing resolution for minimum-ionizing-particles (mips) through the internal amplification of the intrinsic ionization signal through impact-ionization gain in the silicon bulk. Typically, an LGAD includes a highly-doped region just below the junction of a silicon-diode sensor, producing strong electric fields (with peak value of around \SI{3e5}{\volt\per\cm}) that induces controlled electron impact ionization as the charge is collected. This signal amplification enhances the signal-to-noise ratio, significantly reducing the contribution of readout noise to the timing measurement and allowing for the use of thin sensors with rapid charge collection. In this way, LGAD sensors have achieved timing resolution of \SI{17}{\pico\s} for mip timing measurements \cite{CARTIGLIA201783} and frame rates of 500 MHz~\cite{GALLOWAY20195}.

\begin{figure}[htbp]
    \centering
    \includegraphics[scale=0.9]{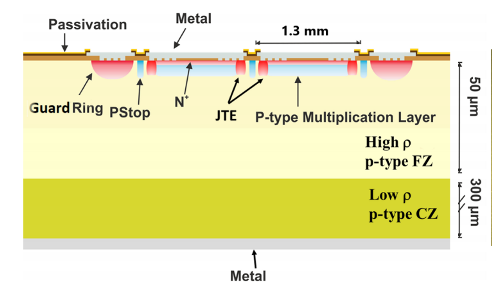}
    \caption{Schematic design of the LGAD Junction Termination Extension (JTE)~\cite{JTE_RD50}.}
    \label{fig:jte}
\end{figure}

However, when such an LGAD sensor is spatially segmented, the strong electric field in the multiplication layer will cause an early breakdown at the edge of the readout electrode if not terminated properly.
To prevent this early breakdown, a ``Junction Termination Extension" (JTE) structure, shown in Figure~\ref{fig:jte}, is introduced in order to reduce the electric field between the segmented implants. 
However the JTE region, which generally extends for \SI{\sim 100}{\micro\m} between neighboring channels, has no signal amplification or collection and is thus inefficient for detecting incident radiation. 
This limits the granularity of practical sensors to the \SI{1}{\mm} scale, since a finer granularity would dramatically decrease the sensor active area. 
However, to make use of LGAD technology to fully exploit the capabilities of future accelerator facilities -- to accomplish ``4D tracking'' for colliding beam detectors \cite{Sadrozinski_2017} or X-ray imaging at next-generation light sources~\cite{ref:Jeph_Wang} -- granularity of better than \SI{100}{\micro\m} will be required.

To achieve higher granularity by overcoming the limitations of the JTE, several variations of LGAD design have been proposed.
For the ``AC Coupled" LGAD (AC-LGAD) \cite{Apresyan:2020ipp}, the highly doped n++ implant of conventional LGADs is replaced with a more resistive n+ layer, with the electrodes AC coupled to the n+ layer through a thin dielectric layer.
This approach eliminates the need for a JTE structure (except at the very edge of the sensor), and can achieve a nearly 100\% fill factor without complicating the fabrication process.
However, AC-LGADs might not be suitable for applications that involve high repetition rate and high signal density due to the AC coupled signal and the intrinsic charge sharing in the resistive n+ layer. 

Another approach, referred to as the ``Inverse" LGAD (iLGAD) \cite{CURRAS2020162545}, features a gain layer that is fabricated on the opposite side of the sensor relative to that of the conventional LGAD, permitting conventional segmentation techniques to be used at the readout surface, which for the iLGAD is thus on the opposite side of the sensor to the gain layer. 
However, iLGADs have yet to achieve a temporal resolution competitive with conventional LGADs. Addtionally, the approach complicates the fabrication process by requiring wafer processing on both sides. 
This process also makes it difficult to produce sensors with an active bulk as thin as the 20-50 $\mu$m thickness characteristic of LGADs.

Recently, a third approach to achieving fine granularity, referred to as the ``Trench-Isolated" LGAD (TI-LGAD) \cite{9081916}, has been proposed that inserts narrow (several micron wide) physical trenches filled with insulator at the boundary of each pixel. For this approach, R\&D is still in an early phase and the efficacy of the approach remains unknown.


Here, we introduce a new variant of LGAD sensor design geared towards higher granularity: the ``Deep-Junction" LGAD (DJ-LGAD). As discussed above, the DJ-LGAD design produces the high-field gain layer within a semiconductor junction that is buried several microns below the readout surface, thereby allowing the electric fields in the bulk just below the sensor to be of similar magnitude to that of a conventional silicon diode sensor. 
In this way the readout plane on top of the device can be spatially segmented without the use of the JTE, allowing for granularity on the scale of tens of microns.

In the following sections, two levels of simulation making use of the Sentaurus TCAD software package~\cite{ref:sentaurus} from the Synopsys corporation are presented. In the first of these (Section~3), performed by the Santa Cruz Institute for Particle Physics (SCIPP), the basic idea of the DJ-LGAD was explored, making use of idealized doping profiles and disregarding potential limitations imposed by the fabrication process. For the second of these (Section~4), performed collaboratively between SCIPP and Cactus materials, the basic elements of the fabrication process were incorporated into the simulation, resulting in the design of a practical device. A first, planar (unsegmented) device was subsequently fabricated at the Cactus Materials facility; rudimentary characteristics of this initial prototype are presented in Section~5.

\section{The Deep Junction LGAD (DJ-LGAD) Concept}

The ``Deep-Junction LGAD" (DJ-LGAD), 
described here for the first time, is a new approach to the application of controlled impact-ionization gain within a silicon diode sensor. The term ``deep-junction" arises from the use of a p-n semiconductor junction buried several microns below the surface of the device. The buried junction is formed by abutting thin, highly-doped p+ and n+ layers, with the doping density chosen to create electric fields large enough to generate impact ionization gain in the narrow buried junction region. Additionally, the doping densities chosen for the p+ and n+ layers are balanced so that when the sensor is fully depleted, the electric field outside of the junction region, while large enough to saturate the carrier drift velocity, is significantly less than that require to create impact ionization gain. This preserves the electrostatic stability at the segmented surface of the detector, thus in principle permitting the production of DC-coupled LGADs with arbitrarily fine granularity. 


\begin{figure}[H]
    \centering
   \includegraphics[scale=0.48]{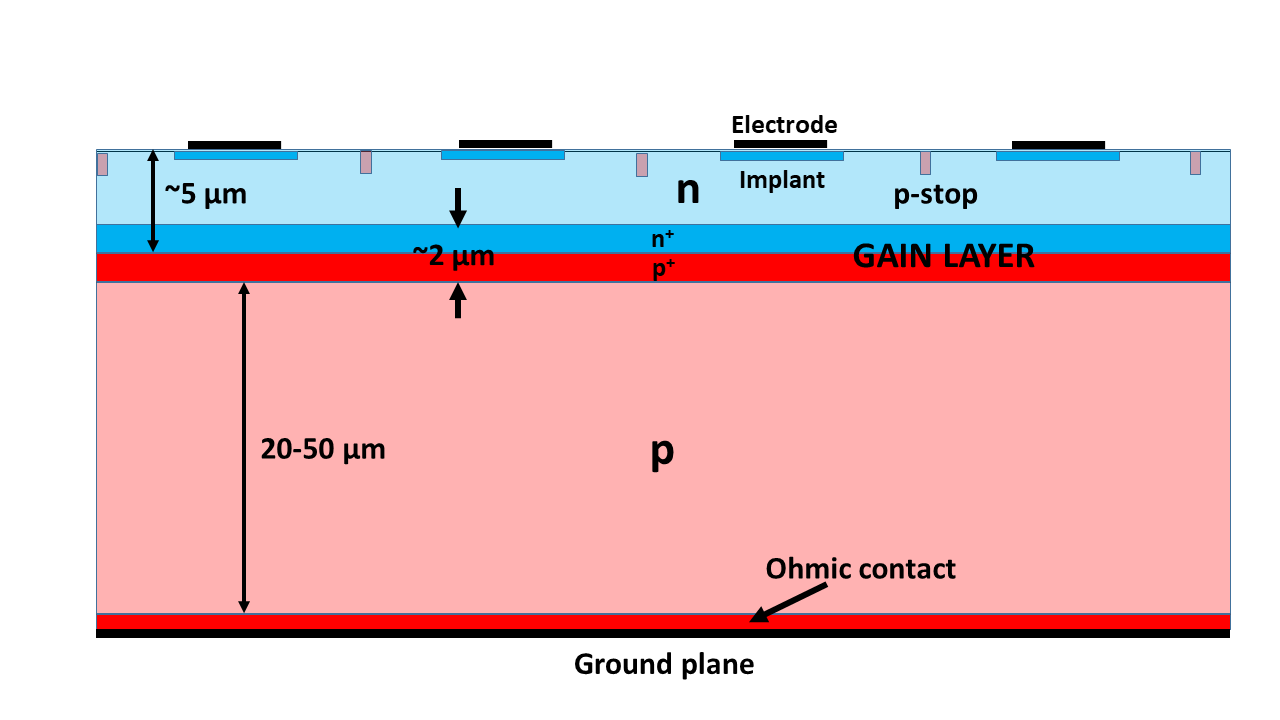}
    \caption{Schematic depiction of the DJ-LGAD concept.}
    \label{fig:design-schematic}
\end{figure}

\begin{center}
\begin{table}[ht]
    \centering
     \begin{tabular}{|| P{4cm} | P{4cm} | P{4cm} ||} 
         \hline
         Element & Doping Level (\SI{}{N/\cm^3}) & Extent In Depth \\ 
         \hline\hline
         N isolation layer (N bulk) & constant doping of \SI{3e12} & From \SI{0}{\micro\m} to beginning of N+ ``gain plate" layer \\  
         \hline
         N+ gain plate (upper half of gain layer) & Gaussian, peak doping \SI{3e16} & peak at \SI{4}{\micro\m}, Gaussian width of \SI{0.17}{\micro\m} \\ 
         \hline
         P+ gain plate(lower half of gain layer) & Gaussian, peak doping \SI{3e16} & Peak at \SI{5.5}{\micro\m}, Gaussian width of \SI{0.17}{\micro\m} \\ 
         \hline
         P drift region (P bulk) & constant doping of \SI{3e12} & End of P+ ``gain plate" layer to \SI{50}{\micro\m} \\ 
         \hline
         P stop & constant doping of \SI{1e13} & \SI{1}{\micro\m} deep at surface, \SI{1}{\micro\m} wide \\ 
         \hline
         N++ implant (under electrode) & constant doping \SI{1e19} & at surface \\ 
         \hline
         Gain layer doping tolerance (N+ and P+ varied together) & effective operation peak doping between \SI{2.9e16} and \SI{3.5e16} & \\ 
         \hline
    \end{tabular}
    \caption{Doping profile assumed for the initial realization of the DJ-LGAD device.}
    \label{tab:sim-spec}
\end{table}
\end{center}

The bulk of the sensor is formed from high-resistivity n-type and p-type silicon. Above the buried junction, between the junction and the segmented readout plane, a several-micron-thick layer of high-resistivity n-type material (the ``isolation layer") is used so that, upon depletion, it contributes very little fixed charge in the isolation layer, leading to a relatively uniform, moderate electric field in the region between the junction and readout plane. Below the junction -- between the junction and the planar biasing electrode -- a thicker layer of high-resistivity p-type material serves as the charge generation medium for through-going charged particles (mips) or absorbed X-rays or heavy ions. A schematic of the proposed doping profile for an initial DJ-LGAD design, showing the bulk, readout, and junction regions described above, is provided in \Fig{\ref{fig:design-schematic}}.

Making use of idealized doping profiles, the Sentaurus simulation package~\cite{ref:sentaurus} was used to explore the possibility of developing such a device, and to specify the doping levels needed to satisfy the design goals stated above. The doping profile of a model device that arose from these studies, featuring a 50 $\mu$m thickness and a transverse segmentation (``channel") pitch of 20 $\mu$m, is displayed in Table~\ref{tab:sim-spec}, which provides a numerical summary of the doping characteristics assumed in the various regions of the simulated sensor. 

\begin{figure}[htbp]
    \centering
    \includegraphics[scale=0.35]{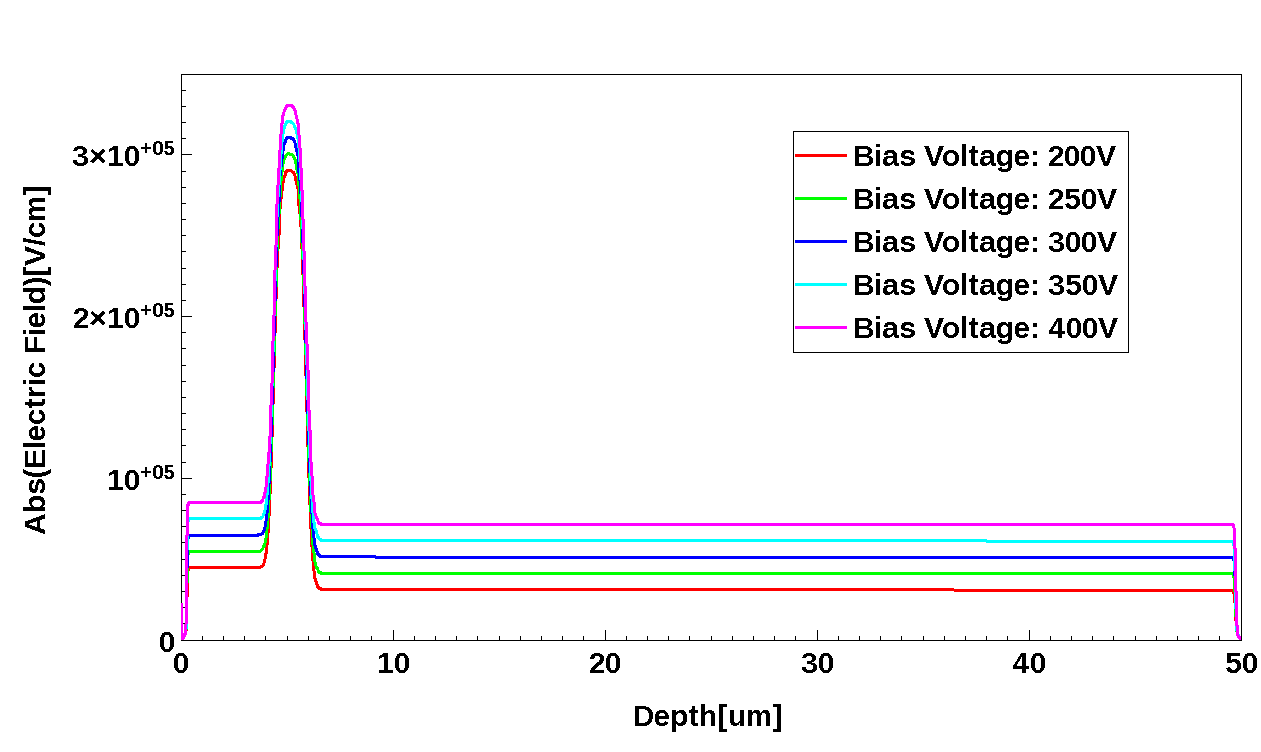}
    \caption{Electric field strengths along a meridian of the DJ-LGAD device described by Table~\ref{tab:sim-spec} that passes through the center of a channel, over a range of applied reverse bias voltages.}
    \label{fig:field-bias}
\end{figure}

The magnitude of the electrostatic field for the doping profile of Table~\ref{tab:sim-spec}, along a longitudinal meridian through the center of a channel, is shown as a function of applied bias voltage in Figure~\ref{fig:field-bias}. For all displayed bias voltages, the \SI{50}{\micro\meter} thick sensor is fully depleted. The peak in the electric field strength arises in the highly-doped buried junction region, and reaches values large enough to induce impact ionization gain for free electron carriers. The field strength in the regions outside of the gain region is large enough to saturate the drift velocity, but remains small enough to protect the device from breakdown between neighboring channels, without the inclusion of a JTE structure. Since the mean-free path for impact ionization depends sensitively on the electric field, the device gain can be adjusted through control of the bias voltage, without compromising the functionality in the bulk and electrode regions. 

\begin{figure}[htbp]
    \centering
    \includegraphics[scale=0.3]{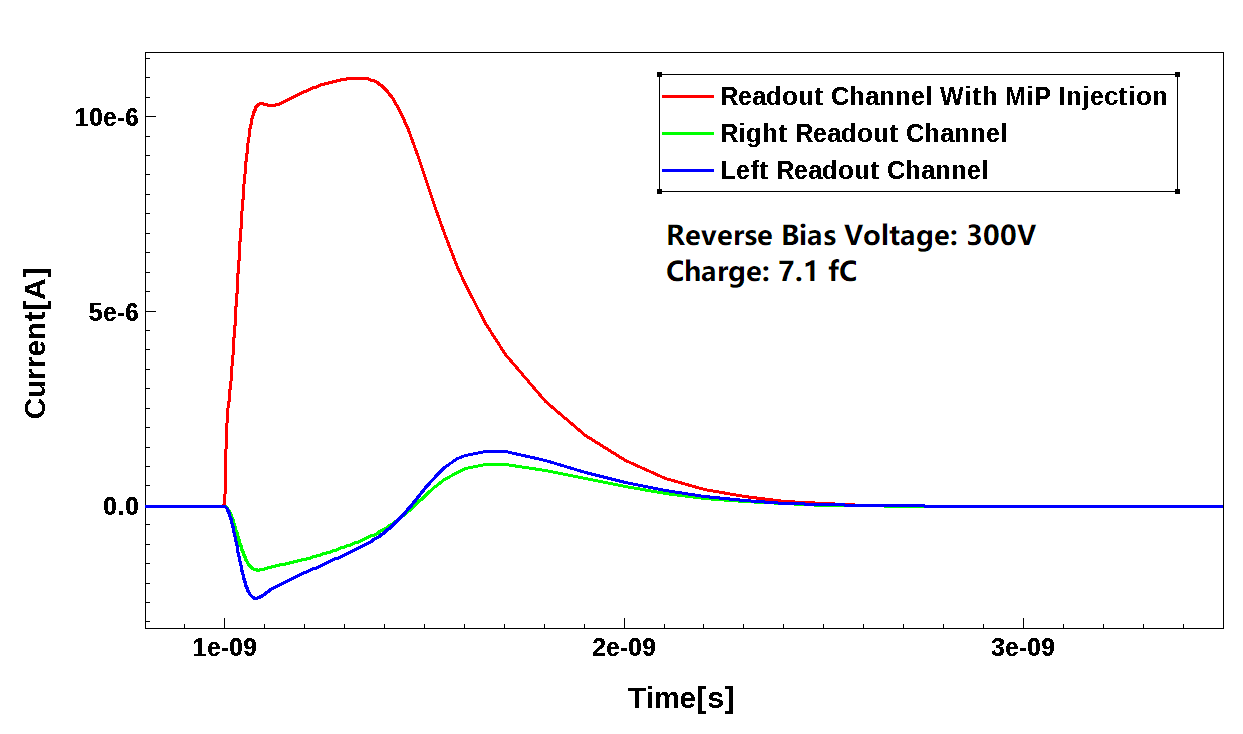}
    \caption{Simulated signal arising from a mip injection along a meridian of the DJ-LGAD device described by Table~\ref{tab:sim-spec} that passes through the center of a channel. The red trace is from the readout channel collecting the mip charge, and the blue and green traces are from the two neighboring readout channels.}
    \label{fig:array-sig}
\end{figure}

In order to study the signal response and gain of this DJ-LGAD model, the heavy ion feature of Sentaurus is used to simulate a mip passing through the device. A uniform energy transfer of \SI{0.0128}{\femto\coulomb\per\micro\m}, typical for mip energy deposition in thin silicon, is defined in a vertical line between the upper and lower surfaces of the device in order to simulate the deposited energy from a mip injection. \Fig{\ref{fig:array-sig}}, for a bias voltage of 300 V, shows the signal response from a mip injected at the center of a channel. The electrode of the channel for which the mip is injected experiences a non-zero integral response with a rise time of order \SI{100}{\pico\second}, while nearby electrodes have signals of significantly smaller magnitude, initially opposite in polarity but then integrating to zero within \SI{1}{\nano\second}. 

\begin{figure}[htbp]
    \center
    \includegraphics[scale=0.6]{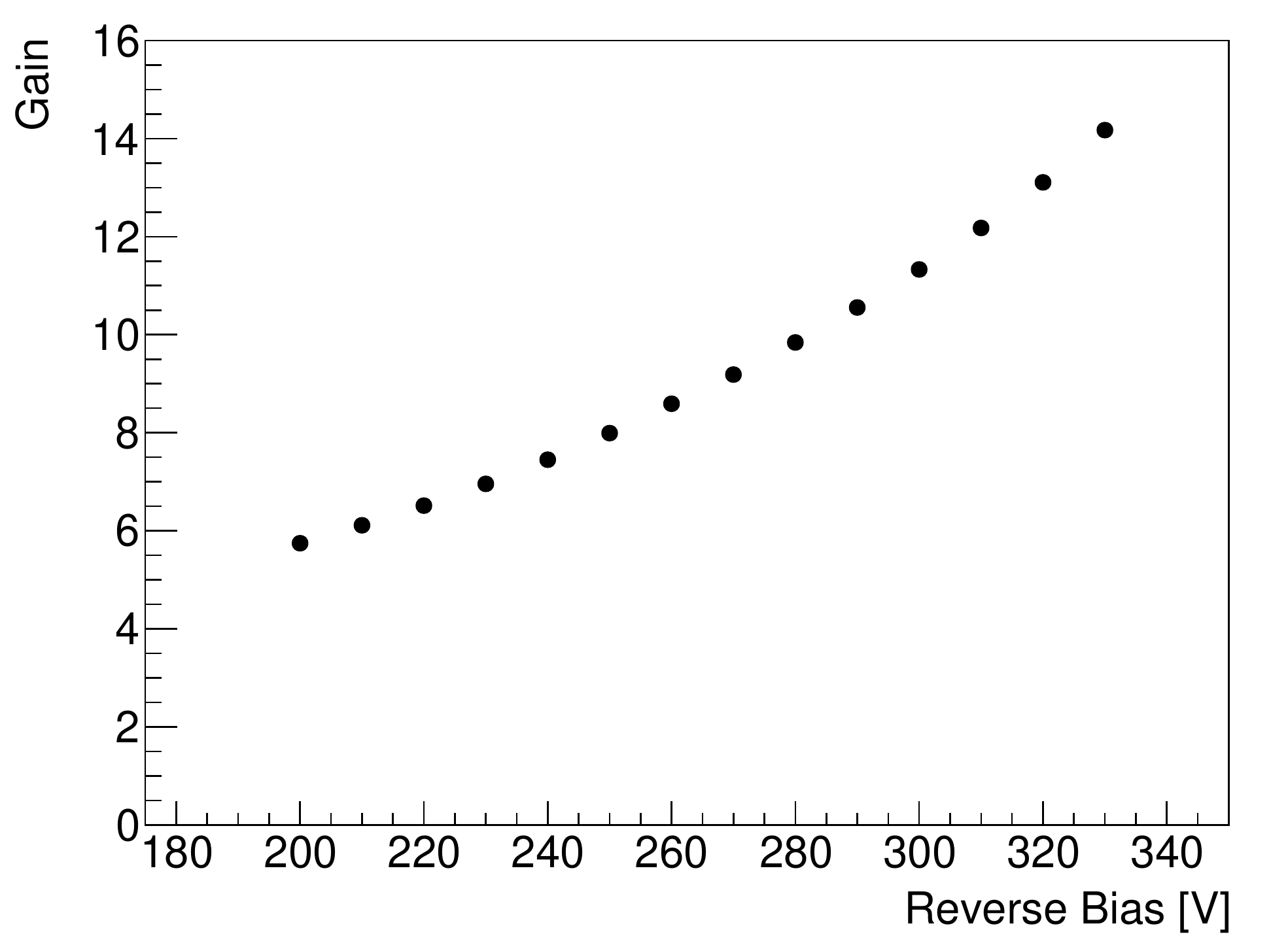}
    \caption{Simulated DJ-LGAD gain as a function of applied reverse bias voltage for the conceptual version of the DJ-LGAD, as described in Table~\ref{tab:sim-spec}.}
    \label{fig:gain-bias}
\end{figure}

To characterize the gain of the device, the mip injection simulation is performed on a simulated reference diode (pin diode) with the same structure of DJ-LGAD but without the gain layer, and the gain is determined as ratio of the integrated signal charge of the DJ-LGAD relative to that of the pin diode. As seen in \Fig{\ref{fig:gain-bias}}, impact ionization gain in excess of 10 is achieved for bias voltages greater than 280 V, with the gain varying smoothly with applied bias voltage.

\begin{figure}[H]
    \centering
    \includegraphics[scale=0.6]{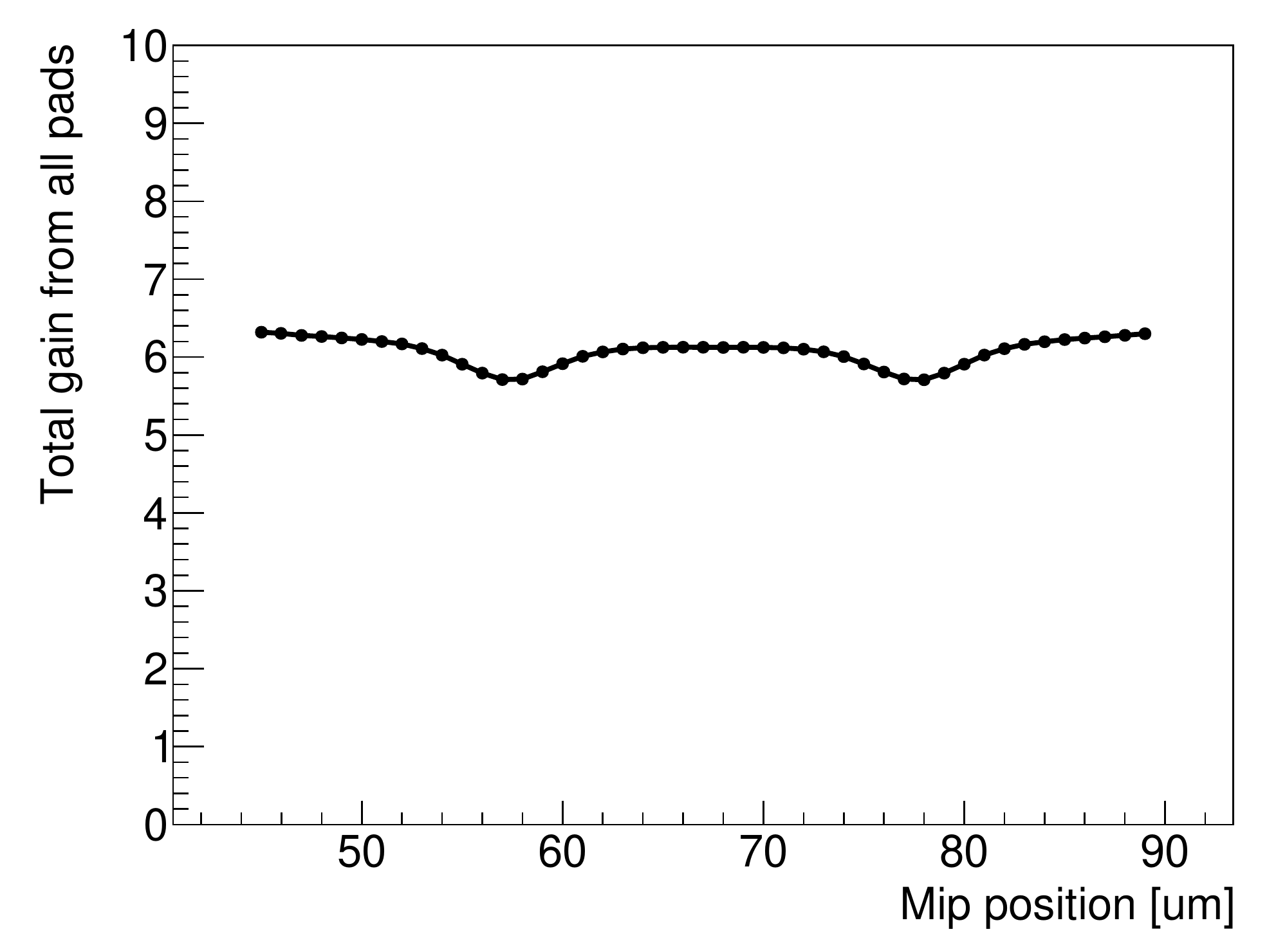}
    \caption{Sum of the integrated signal charge over all channels, as a function of the transverse position of incidence of a simulated mip, for the 20 $\mu m$ pitch sensor of Table~\ref{tab:sim-spec}, biased at approximately 200 V. Uniformity within $\pm$5\% is observed}
    \label{fig:mip-scan}
\end{figure}

As discussed in Section~2, the primary advantage of the DJ-LGAD approach is the elimination of the JTE structure, which creates ``dead regions" of limited detector response over many tens of microns between neighboring channels. In contrast, no JTE structure is required for a DJ-LGAD array, as the buried gain layer ensures a uniform gain performance across channels. The resulting gain uniformity can be studied by exploring the simulated sensor response as a function of the transverse location of the mip injection, with the device held at constant bias voltage.
\Fig{\ref{fig:mip-scan}} shows the DJ-LGAD response uniformity for an aggressive channel pitch of 20 $\mu$m, for the device of Table~\ref{tab:sim-spec} biased to approximately 200 V. To account for the sharing of the signal when a mip is injected in the region midway between the center of two channels, the sum of gain from all enabled channels is formed for each transverse location of the mip injection. The variation in the gain is seen to be within $\pm 5$\% across the device, with the loss in gain in the region between channels caused by a small distortion of the gain-layer fields that is introduced by the narrow inter-channel p-stop insertion. At this level of simulation, the DJ-LGAD approach is seen to hold significant promise towards the development of a highly-pixellated DC-coupled silicon diode sensor with substantial internal gain and precise temporal resolution.

\section{Practical Realization of the DJ-LGAD Device}

As a first realization of the DJ-LGAD concept, several closely-related versions of a planar (non-segmented) device have been designed and fabricated, based on a wafer-to-wafer bonding approach to establish the buried junction interface. The goals of this effort were to demonstrate the feasibility of the deep junction idea, and to confirm several details of the design and fabrication process essential to the development of the device, including the implantation and bonding processes, and the junction termination and guard-ring structures at the edge of the active area. In moving from an idealized doping profile to one arising from the simulation of the fabrication process, the performance characteristics might be expected to change somewhat from those of the conceptual devices described in the previous section. Thus the basic performance characteristics are revisited in this section, after the description of the augmented simulation and optimization procedure. 

In the wafer-to-wafer bonding approach, ion implantation is used to establish the highly-doped n+ and p+ regions, respectively, near the surface of two high-resistivity (1,0,0) wafers. The surfaces of these two wafers are abutted against one another, and the interface barrier arising from surface roughness and other imperfections is then removed by high-temperature annealing, leading to a clean surface-to-surface bond through which charge (electrons and holes) can flow freely. 

The ion implantation energy for the p-type dopant (boron) was set to 735 keV, the maximum energy achievable by the implantation apparatus. To avoid excessive ion channeling along crystal plane boundaries, the wafers are canted at an angle of \SI{7}{\degree} relative to the normal axis of the wafer during the implantation process. This leads to a most-likely ion penetration depth, and a corresponding peak in the doping profile, at a depth of approximately 0.9 $\mu$m. 

\begin{figure}[H]
    \centering
      \includegraphics[width=1\linewidth]{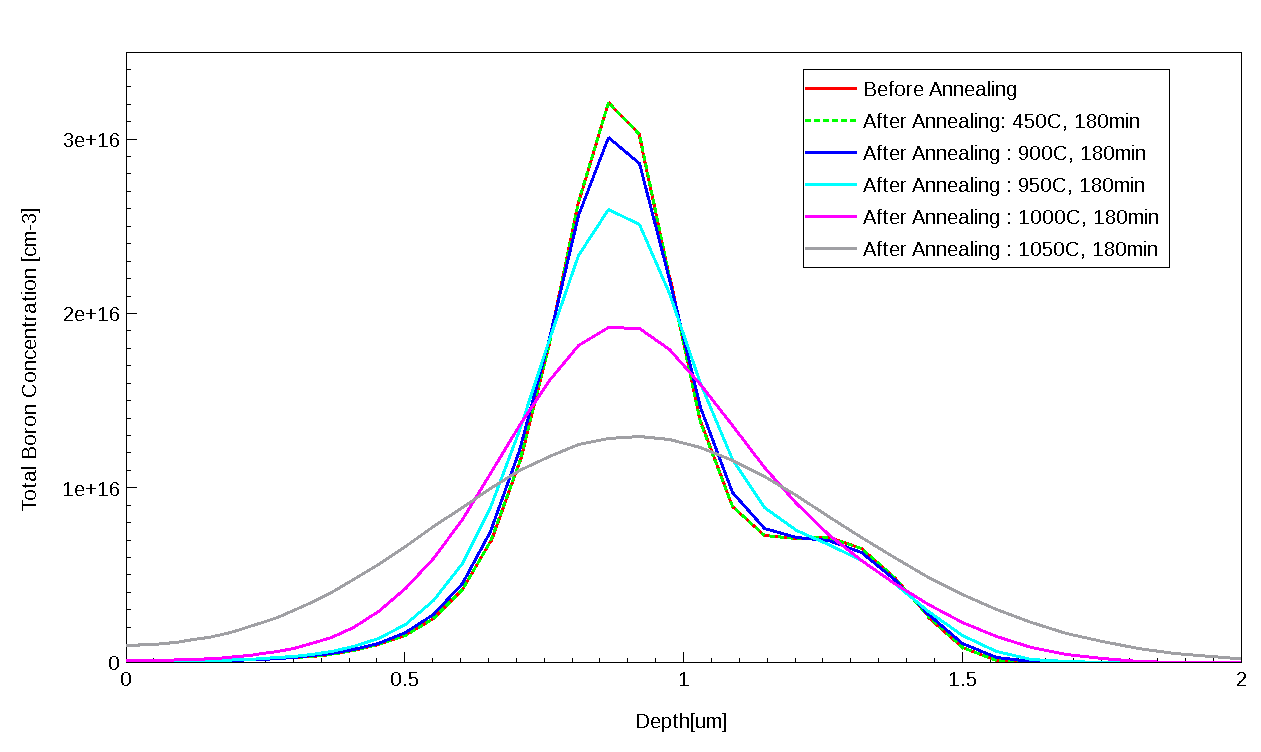}
      \centering
      \includegraphics[width=1\linewidth]{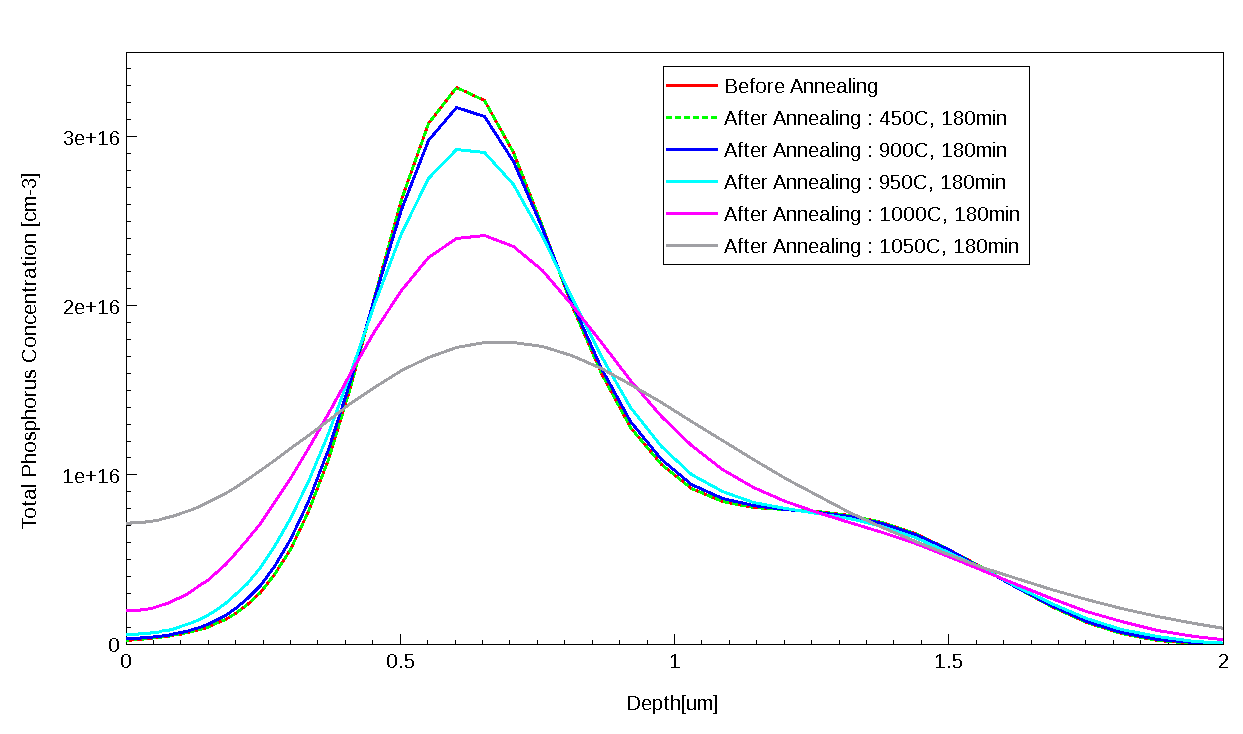}
    \caption{Gain layer doping profile as a function of annealing temperature (180 minute annealing episode)}
    \label{fig:annea60ng}
\end{figure}

To establish the 1.5 $\mu$m separation between the peaks of the n+ and p+ doping profiles (see Table~\ref{tab:sim-spec}), the implantation energy of the n-type dopant (phosphorus) was then tuned within the Sentaurus simulation to produce a doping profile with a peak at a depth of 0.6 $\mu$m (again for a cant of \SI{7}{\degree}), corresponding to an implantation energy of 450 keV. Simulation of the resulting n-type and p-type doping profiles, as a function of annealing temperature (180 minute annealing episode) is shown in Figure~\ref{fig:annea60ng}. For the simulation in this figure, an ion-beam fluence of $1.60 \times 10^{12}$ ($1.52 \times 10^{12}$) particles per cm$^2$ was assumed for the n-type (p-type) implantation process. While the simulation suggests that some channeling effects remain that produce tails in the depth profile, the bulk of the distribution is adequately peaked at the desired depth. The simulation suggests that modifications to the doping profile during the annealing process are expected to be minimal for temperatures below \SI{950}{\degree} C.   

\begin{figure}[htbp]
    \centering
    \includegraphics[scale=0.48]{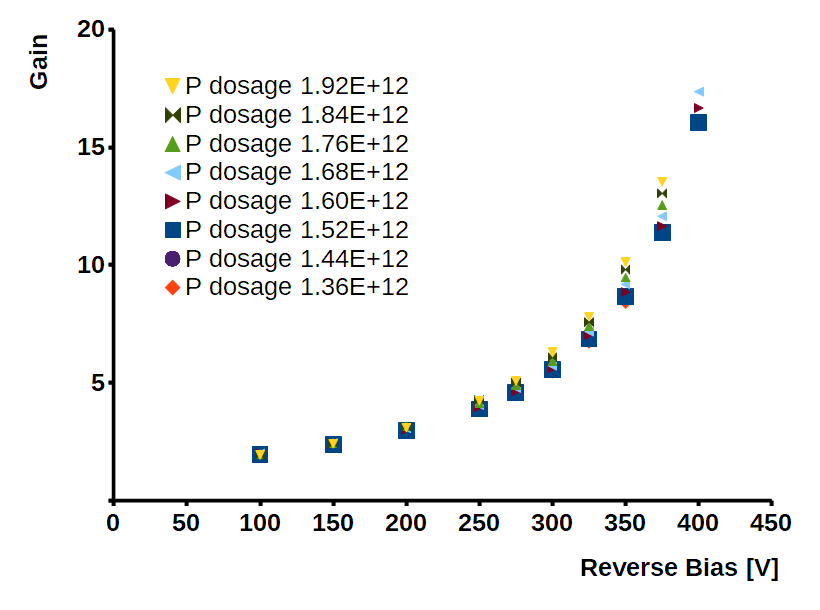}
    \caption{Deep-junction gain curve for various p+ implantation ion-beam fluences, assuming an n+ implantation fluence of $1.60 \times 10^{12}$.}
    \label{fig:manygain}
\end{figure}

The impact-ionization gain of a device making use of these implantation-beam energies was simulated for a buried-junction depth of 5 $\mu$m, and was found to be sufficiently insensitive to the precise value of the implantation beam fluence. For example, Figure~\ref{fig:manygain} shows the impact-ionization gain as a function of bias voltage for a range of boron ion-beam fluences, assuming a phosphorus ion-beam fluence of $1.60 \times 10^{12}$. A total detector thickness of 50 $\mu$m is assumed. The gain is seen to be relatively stable over a range of $\pm$20\% in the corresponding doping level. A beam fluence of $1.52 \times 10^{12}$ ions per cm$^2$ was found to produce the most slowly-varying, and therefore controllable, gain, and was chosen as the p-type implantation doping level for the prototype device. 

\begin{figure}[htbp]
    \centering
    \includegraphics[scale=0.4]{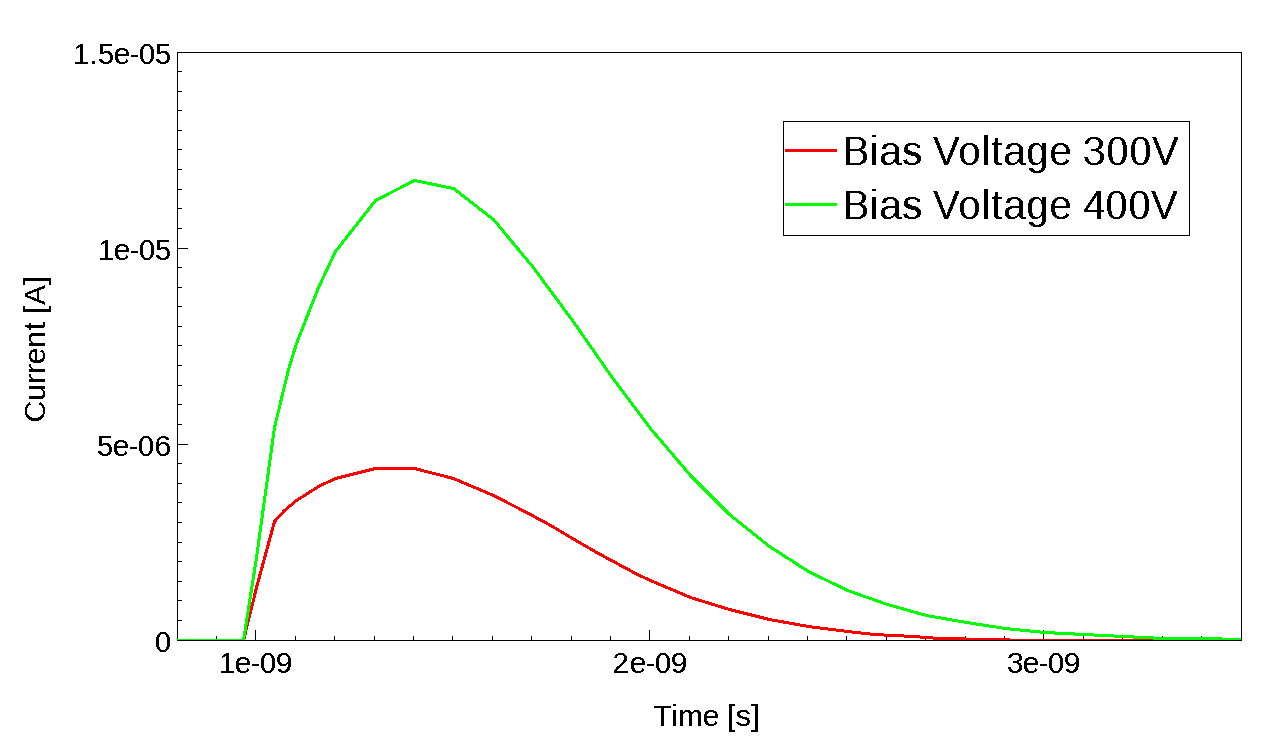}
    \caption{Simulated charge collection response to a mip injection for bonded wafers doped according to the procedures described in the text. The total sensor thickness was set to 50 $\mu$m.}
    \label{fig:current_pulse}
\end{figure}

A simulation of the charge-collection response (current as a function of time), for a planar DJ-LGAD fabricated as described above, is displayed in Figure~\ref{fig:current_pulse}. The pulse is shown for two values of the applied bias voltage (300 V and 400 V), for which the impact-ionization gain factor is approximately 5 and 15, respectively. For this more realistic model of the DJ-LGAD, a rise time of approximately 250 psec is observed, consistent with that observed for conventional surface-junction LGADs.

\begin{figure}[H]
    \centering
        \includegraphics[width=1\linewidth]{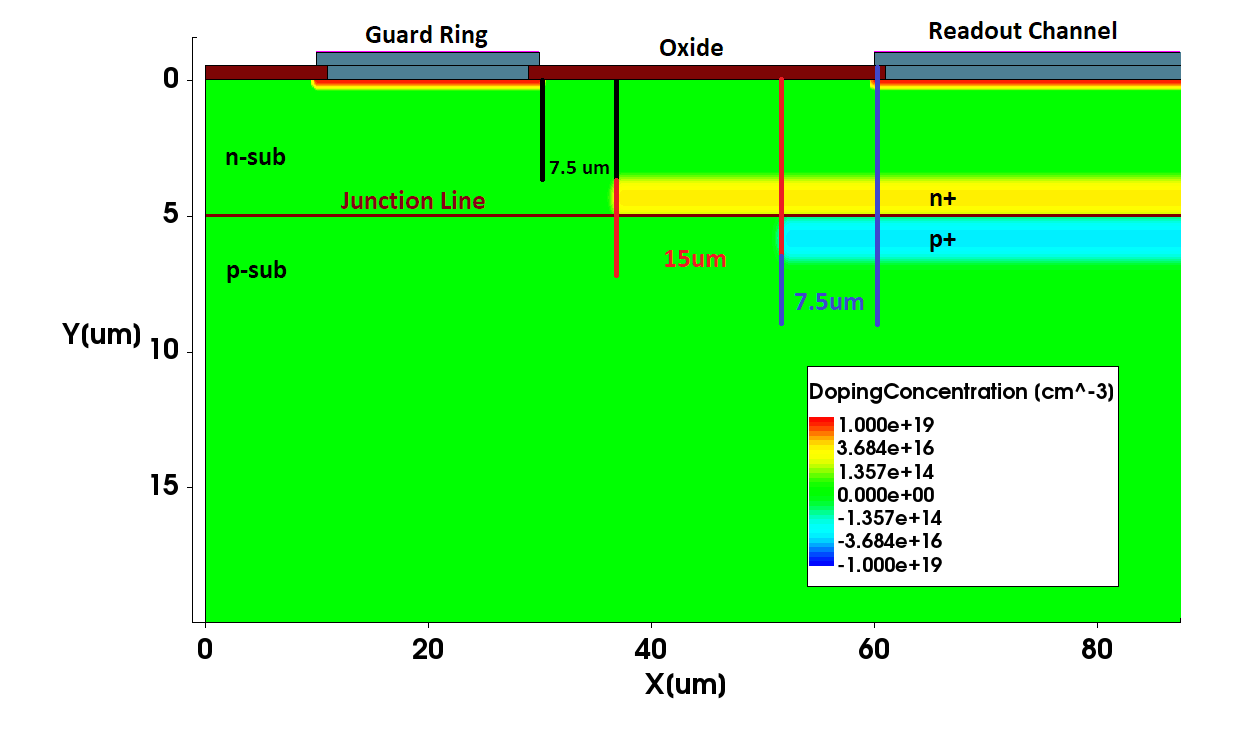}
        \includegraphics[width=1\linewidth]{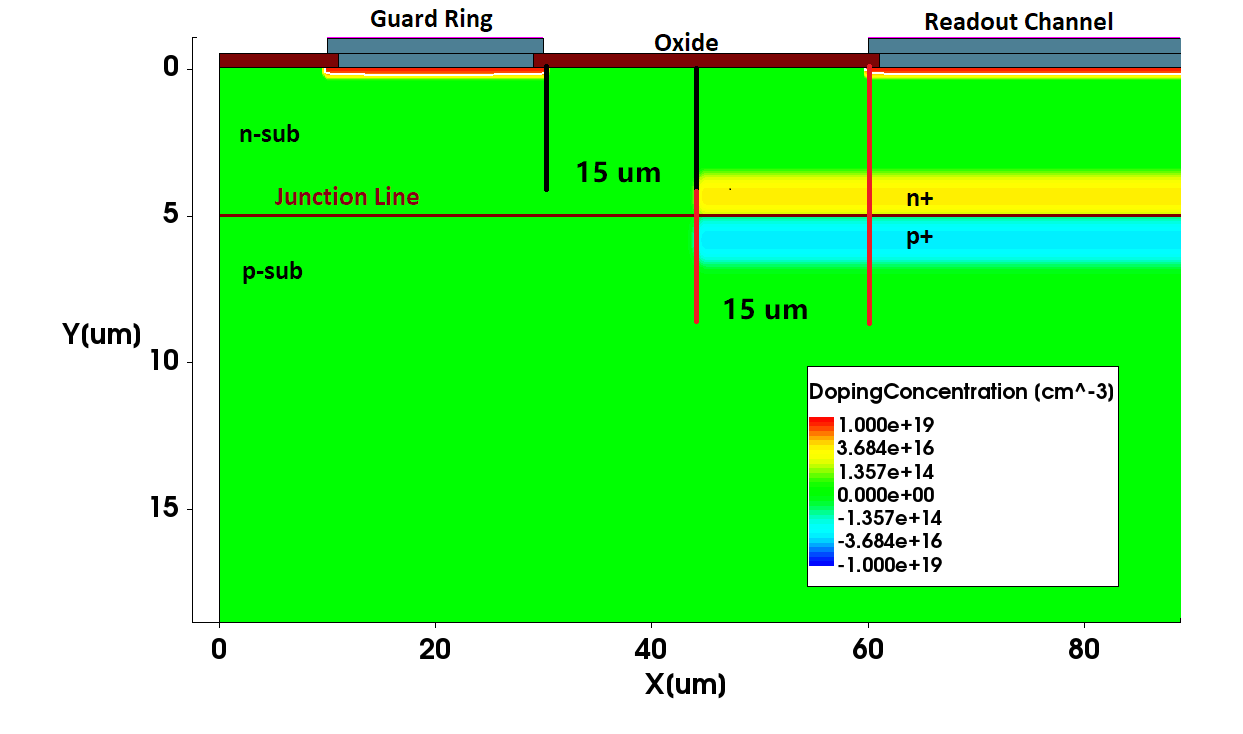}
    \caption{Doping profile cross section for the asymmetric termination(top) and symmetric termination(bottom) scheme. For the symmetric termination scheme, the n+ and p+ regions terminate at the same point in the $x$ coordinate, half-way between (15 $\mu$m from) the right edge of the guard ring and left edge of the readout pad.}
    \label{fig:device_param_naming}
\end{figure}

In producing an operable device, it is essential to properly terminate the high-field semiconductor junction at the end of the active area in a manner that avoids early (low bias voltage) breakdown of the sensor. Simulation studies suggested that this can be achieved by terminating the n+/p+ semiconductor junction in the region immediately below the gap between the planar readout electrode and a conductive guard ring encircling the readout electrode at a distance of 30 $\mu$m. Two types of junction termination are considered: one for which the upper (n+) and lower (p+) implants terminate at the same point (``symmetric termination") and one for which the upper implant extends 15 $\mu$m further towards the guard ring than does the lower implant (``asymmetric termination"). Figure~\ref{fig:device_param_naming} shows the two-dimensional doping profile of the asymmetric termination scheme; for the symmetric scheme, the n+ and p+ regions terminate at the same transverse position, half-way between the edges of the readout electrode and guard ring. 

\begin{figure}[H]
    \centering
      \includegraphics[width=1\linewidth]{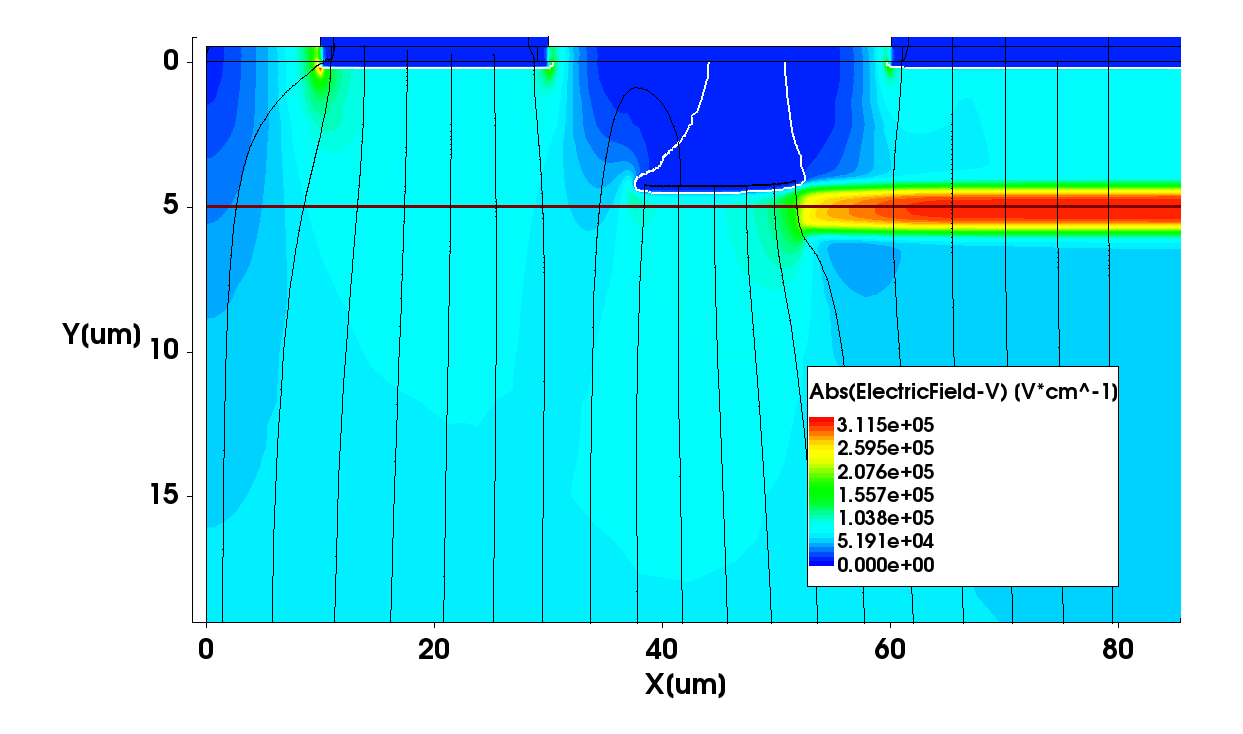}
      \includegraphics[width=1\linewidth]{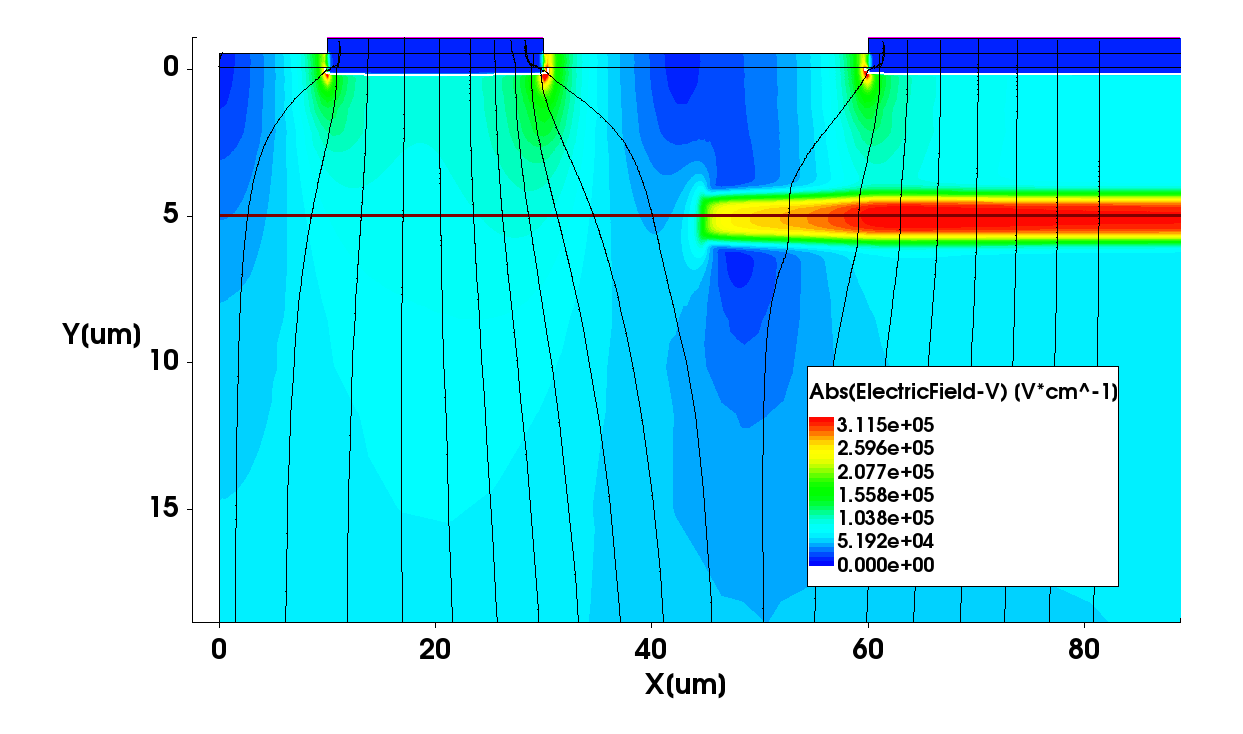}
    \caption{Electric field map in the region of the gain layer termination, assuming a bias voltage of 300 V and a sensor thickness of 50 $\mu$m. The upper (lower) plot shows the expected electric field for the asymmetric (symmetric) termination scheme. The electric field lines are shown in black, while the magnitude of the electric field is indicated by the color. For the upper plot (asymmetric termination), the region enclosed by the thin white contour is undepleted. However, this narrow undepleted region lies outside the active area of the sensor, and thus does not degrade the performance of the device relative to that of the device with symmetric termination.}
    \label{fig:termination_field_300V}
\end{figure}

\begin{figure}[H]
    \centering
      \includegraphics[width=1\linewidth]{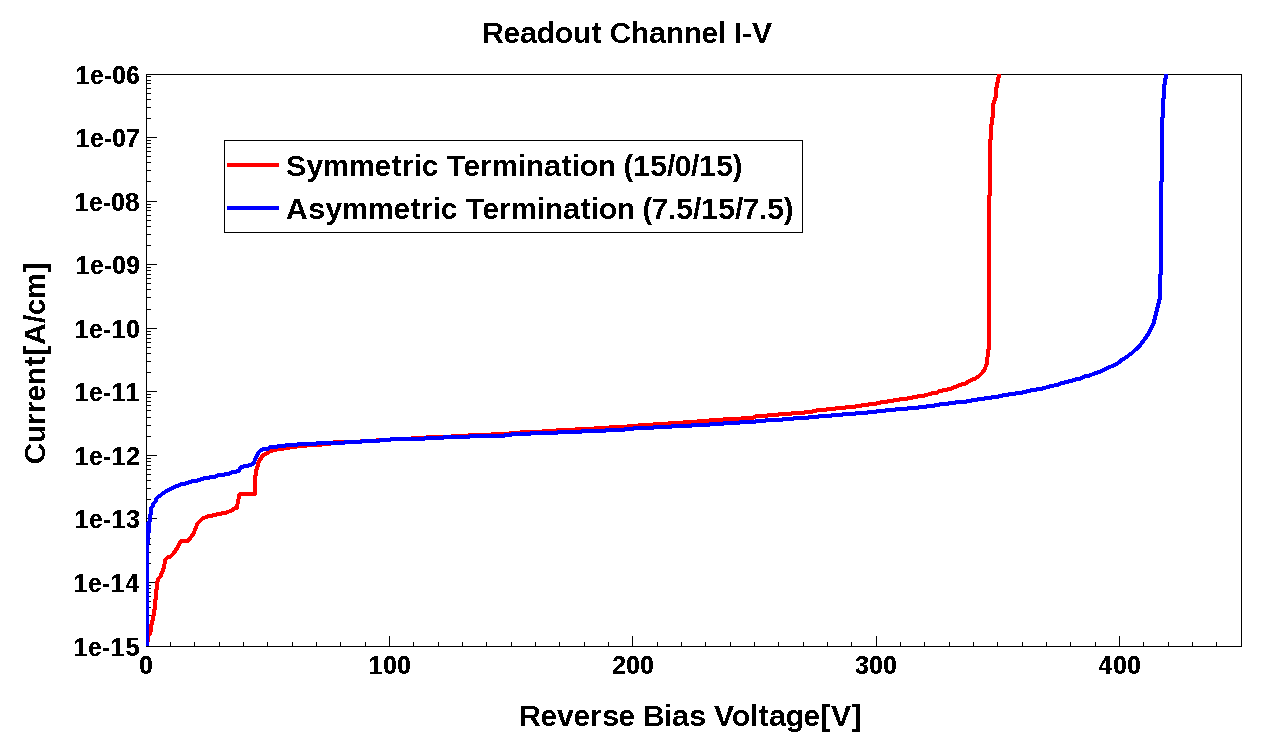}
      \includegraphics[width=1\linewidth]{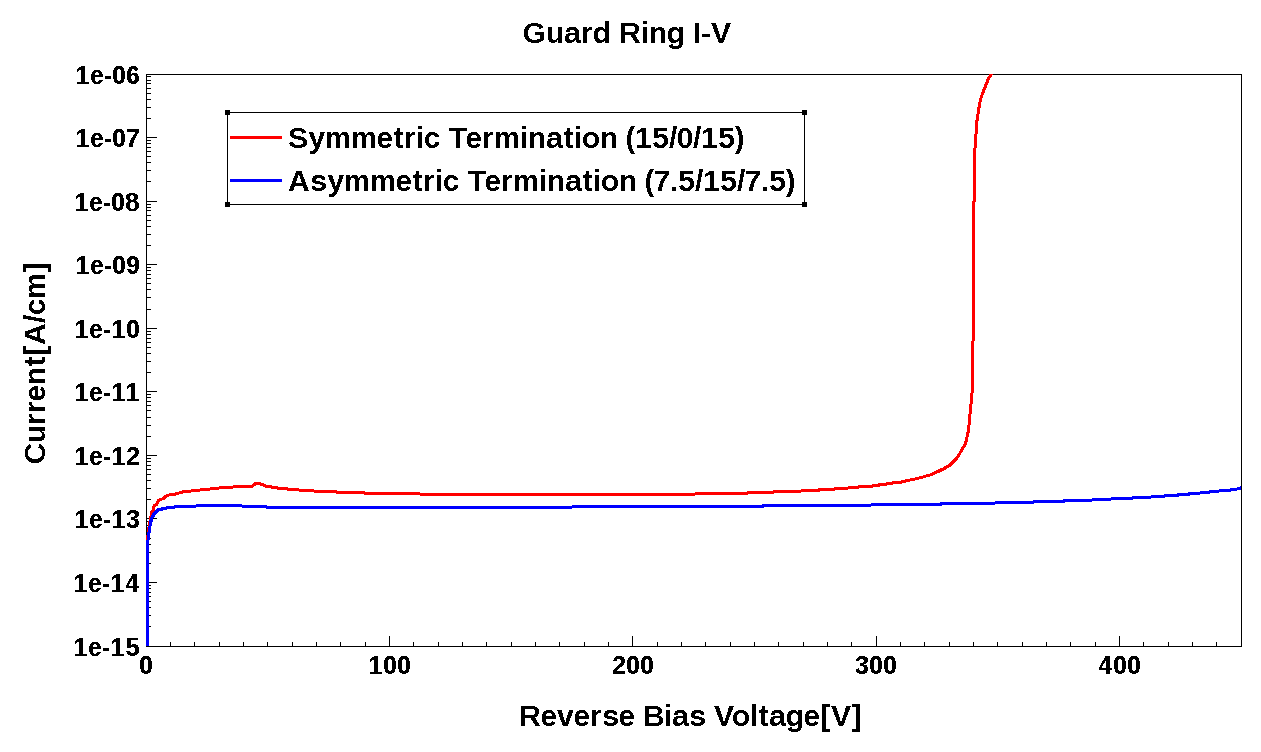}
    \caption{Expected dependence of the linear leakage current density vs reverse-bias voltage (I-V curve) for the readout electrode (upper) and guard ring (lower). The expectation for the symmetric termination scheme is shown in red, while that for the asymmetric scheme is shown in blue.}
    \label{fig:term_IV}
\end{figure}

The electric field configuration expected for these symmetric and asymmetric doping profiles, assuming an applied reverse bias of 300 V and an overall detector thickness of 50 $\mu$m, is shown in the region of the junction termination in Figure~\ref{fig:termination_field_300V}. The asymmetric scheme is seen to provide a lower and more spatially uniform field, especially in the region near the electrode boundaries. The corresponding expectations for the dependence of leakage current upon bias voltage is shown in Figure~\ref{fig:term_IV}, for both the guard ring and readout electrodes. While both termination configurations are expected to allow for operation well above 300 V before breakdown, as anticipated from Figure~\ref{fig:termination_field_300V}, the asymmetric termination scheme is expected to provide a significantly greater operating range than that of the symmetric scheme. A final junction-termination study showed no difference in the detector breakdown properties when a 1-2 $\mu$m thick p-spray structure with a doping density of approximately $5 \times 10^{13}$ cm$^{-2}$ was embedded in the surface of the device, interposing between the readout electrode and guard ring.

\section{Initial Characterization of the Fabricated Device}
\begin{itemize}
\item Whatever we can have - first say what configurations and splits were fabricated. Was p-apray used as simulated?
\end{itemize}

\section{Acknowledgements}
This work was supported by ... WRITE SBIR NUMBER.
This work was supported by the United States Department of Energy, grant DE-FG02-04ER41286, and partially performed within the CERN RD50 collaboration. We also acknowledge helpful conversations with and consideration by SCIPP faculty member Hartmut F.-W. Sadrozinski.

\bibliography{bib/DJ_LGAD,bib/TechnicalProposal,bib/hpk_fbk_paper,bib/HGTD_TDR}

\end{document}